\documentclass[aps,twocolumn,graphics,floatfix,tightenlines]{revtex4-2}
\usepackage{graphics}
\usepackage{epstopdf}
\usepackage{epsfig}
\usepackage{graphicx}
\usepackage{epsf,epic}
\usepackage{color}
\usepackage{subfig}
\usepackage{mathtools}
\usepackage{amsmath}
\usepackage{booktabs}
\usepackage{multirow}
\usepackage{physics}
\usepackage{hyperref}
\usepackage{amsfonts}
\usepackage{wrapfig}
\usepackage{breqn}
\usepackage{pstricks}
\usepackage[utf8x]{inputenc}
\usepackage{multirow}
\usepackage{fancyref}
\usepackage{pst-node}
\usepackage{soul}
\usepackage{float}
\usepackage{bm}
\usepackage{dcolumn}
\newcommand{\etal}{\textit{et al.\ }}
\newcommand{\ie}{\textit{i.e.\ }}

\makeatletter
\usepackage{etoolbox} % for \appto
\appto{\appendix}{%
	\@ifstar{\def\theequation@prefix{A.}}%
	{}%
}
\preto\maketitle{%
	\begingroup\lccode`~=`,
	\lowercase{\endgroup
		\let\saved@breqn@active@comma~% save breqn active comma
		\let~}\active@comma % set the active comma to what revtex4-1 wants
}
\appto\maketitle{%
	\begingroup\lccode`~=`,
	\lowercase{\endgroup
		\let~}\saved@breqn@active@comma % undo the change
}
\makeatother
\begin{document}
	
	\title{Electronic band structure and exciton properties of $Pna2_1$ CaSnN$_2$}
	
	\author{Ilteris K. Turan}
	\email{ilteris.turan@case.edu}
	\author{Sarker Md. Sadman}
	\email{sarker.sadman@case.edu}
	\author{Walter R L. Lambrecht}
	\email{walter.lambrecht@case.edu}
	\affiliation{Department of Physics, Case Western Reserve University, 10900 Euclid Avenue, Cleveland, OH 44106-7079, USA}
	
\begin{abstract}
  The electronic band structure of CaSnN$_2$ in the wurtzite based $Pna2_1$ structure is calculated using the Quasiparticle Self-consistent (QS)$GW^{\rm BSE}$ method including ladder diagrams in the screened Coulomb interaction $W^{\rm BSE}$ and is found to have a direct gap of 2.59 eV at $\Gamma$, which corresponds to blue light wavelength of 478 nm and makes it an attractive candidate for sustainable blue light-emitting diodes (LEDs), avoiding Ga and In. The valence band splitting is analyzed in terms of symmetry labeling and the effective mass tensor is calculated for several bands at $\Gamma$. 
  The valence band maximum has $a_1$ symmetry and gives allowed transitions to the conduction band minimum (also of $a_1$  symmetry) for light polarized along the {\bf c}-direction. While this is unfavorable for light emission with transverse electric (TE) or s-polarization from the basal plane, this would not be an impediment if another surface than the basal plane is used.  Furthermore 
  the crystal field splitting between the $a_1$  and $b_1$ states, corresponding to polarizations along {\bf c} and {\bf a} respectively, reverses under an applied uniaxial tensile strain of 3.7\% along the {\bf c} direction, which might occur under biaxial compressive strain in the basal plane. 
  The optical dielectric function including electron-hole interaction effects is also reported and the excitons are analyzed, including several dark excitons.  	
\end{abstract}
\maketitle
	
\section{Introduction}
The importance of the group-III nitrides, including AlN, GaN, InN and their alloys and hetero-structures, which form the basis of white light-emitting diodes and blue lasers, was recognized with the Nobel Prize in Physics of 2014 \cite{nobel2014}.
While work continues to increase the efficiency of these devices and to extend their operation into the deeper UV region, the problem of their sustainability is also gathering attention as both In and Ga become scarce and more expensive as their widespread use in semiconductors increases. The family of ternary II-IV-N$_2$ nitrides provide a potential solution to this problem with naturally abundant elements such as II=Mg, Zn, Ca and group IV=Si, Ge, Sn. While Ge is also less abundant, both Si and Sn are abundant elements \cite{abundance}. 
The II-IV-N$_2$ semiconductors have been reviewed in \cite{Lambrechtbook,Lyu2019,Martinez17} with the emphasis on Zn-based compounds. Recently, the Mg compounds have been investigated for their potential to achieve higher band gaps \cite{Atchara16,Lyu2019mg,Quirk14,Arab2016,Bruls2000,Pramchu17}. Among the Ca based compounds, CaSiN$_2$ has been synthesized as early as 1924 \cite{Wohler1924,Franck1939,Laurent1968} and CaGeN$_2$ synthesis was reported since 1967 \cite{Guyader1967,Maunaye1971}, but besides the structure reports, its properties have not been much investigated. We are only aware of one band structure calculation of CaSiN$_2$ \cite{deBoer2023}. More recently other alkaline earth silicon nitrides have been synthesized \cite{Gal2004}. 
        
CaSnN$_2$ high pressure growth was reported by Kawamura \etal\cite{Kawamura2021}. They found a rock-salt type structure with space group $R\bar{3}m$.
On the other hand, the Materials Project lists CaSnN$_2$ as having a wurtzite based $Pna2_1$ structure in entry mp-1029633 \cite{mp-1029633}, but no information is provided on the electronic structure. Here we report computational predictions of CaSnN$_2$ in the $Pna2_1$ structure, and predict a direct band gap in the blue range of the visible spectrum. CaSnN$_2$ could thus be an attractive replacement for In$_x$Ga$_{1-x}$N in blue LEDs. Of course this will depend on the capability to grow the material in single crystal thin film form and on the capability of both p and n-type doping. In this paper, we focus only on the band structure and crystal structure of $Pna2_1$ CaSnN$_2$. 
        
\section{Computational Methods}
Our calculations use the full-potential linearized muffin-tin orbital method (FP-LMTO) as implemented in the {\sc Questaal} codes, \cite{questaalpaper} which incorporates both Density Functional Theory (DFT) and Many-Body Perturbation Theory (MBPT).
The DFT calculations of the structural stability are carried out in the PBEsol parametrization of the generalized gradient approximation to exchange and correlation \cite{PBEsol}. To make accurate predictions of the band structure and in particular, the band gap we use the Quasiparticle Self-consistent $GW$ (QS$GW$) \cite{MvSQSGWprl,Kotani07} method, which is a variant of Hedin's $GW$ theory \cite{Hedin65} in which $G$ stands for the one-electron Green's function and $W$ for the screened Coulomb interaction. In this quasiparticle self-consistent version, a non-local exchange correlation potential is extracted from the energy dependent $GW$ self-energy $\Sigma(\omega)=iGW$ in each iteration and its off-diagonal elements in the basis set of the $H^0$ Hamiltonian from which this $G$ and $W$ were derived is included in the next iteration until the self-energy no longer changes. This provides a $GW$ result independent of the DFT starting point $H^0$ and Kohn-Sham eigenvalues from the effective potential $H^0$ which equal the real part of the quasiparticle excitation energies. The specific prescription for extracting the exchange correlation correction is
\begin{equation}
\Delta v_{xc}=\frac{1}{2}|\psi_i\rangle{\rm Re}\left[\Sigma_{ij}(\epsilon_i)+\Sigma_{ij}(\epsilon_j)\right]\langle\psi_j|,
\end{equation}
where summation over repeated indices is assumed. A more recent version of QS$GW$ goes beyond the random phase approximation (RPA) in calculating the screened Coulomb interaction $W({\bf q},\omega)$ by including electron-hole interactions or ladder diagrams in the polarization propagator via a Bethe-Salpeter Equation (BSE) approach \cite{Cunningham23}. It has been called QS$G\hat{W}$ or QS$GW^{\mathrm{BSE}}$ to distinguish it from QS$GW^{\mathrm{RPA}}$.

The BSE are also used to investigate the optical properties in the long wavelength ${\bf q}\rightarrow0$ limit \cite{Onida02,Strinati1988,Hanke78}. Here, we use the implementation in the LMTO and product basis set developed by Cunningham \etal \cite{Cunningham18,Cunningham23}.
        
\begin{figure}
	\includegraphics[width=8.5cm]{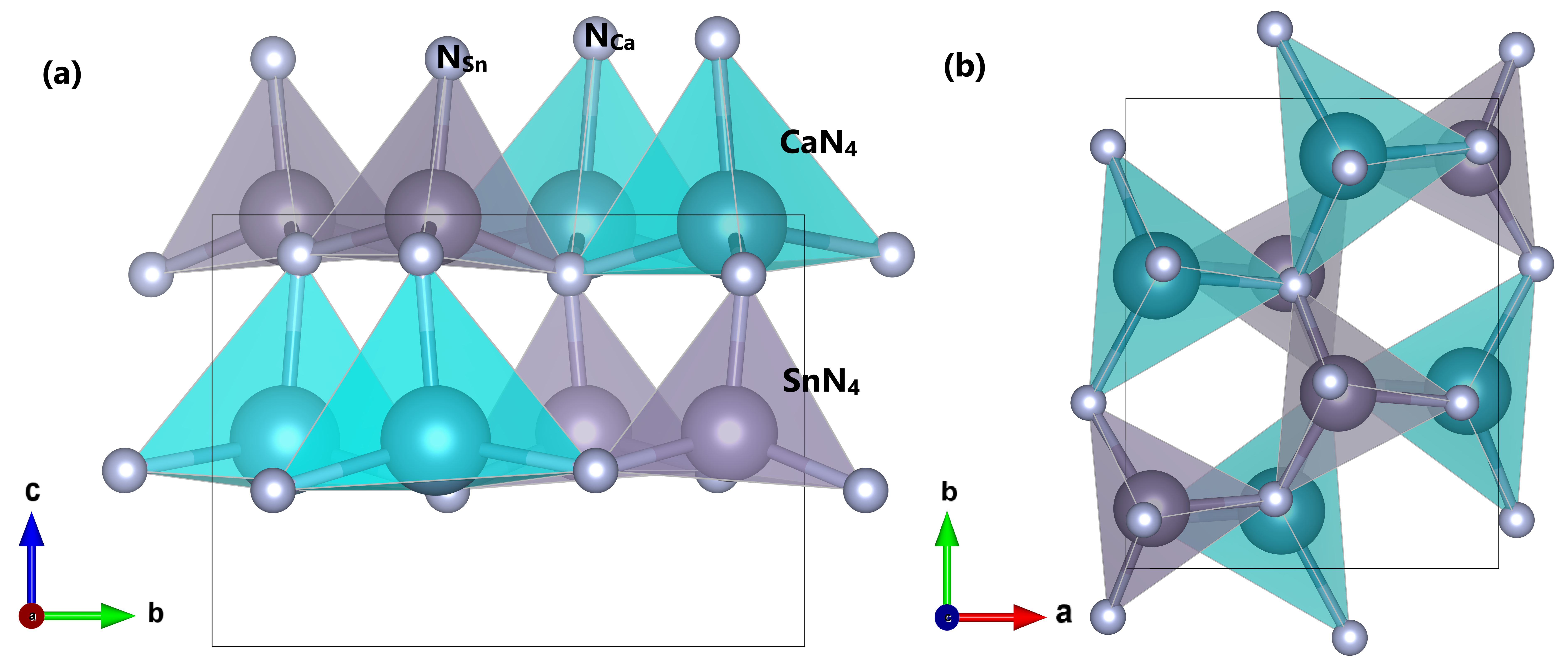}
    \caption{Crystal structure of CaSnN$_2$ with nearest neighbor CaN$_4$ tetrahedra in blue and nearest neighbor SnN$_4$ tetrahedra in gray shown from (a) side view and (b) top view. The image is generated using the VESTA3 software \cite{Vesta}.
    \label{structure}}
\end{figure}

We use a $spdfspd$ envelope basis set on Ca, S and N with augmentation up to $l=4$. The Ca-$3p$ and Sn-$4d$ semi-core levels are treated as local orbitals. The self-energy matrix is approximated by a diagonal average value above a cut-off of 2.15 Ry. A $5\times5\times5$ {\bf k}-mesh is used to sample the Brillouin zone both for charge self-consistent calculations and to sample the self-energy. The direct gap found at the $\Gamma$ point with a $5\times5\times5$ {\bf k}-mesh differs from that of a $3\times3\times3$ mesh by only 0.03 eV within the QS$GW^{\rm RPA}$. For the QS$GW^{\rm BSE}$, the smaller {\bf k}-mesh allows us to include a higher number of valence and conduction bands, for which we take 24 valence bands and 12 conduction bands. This includes all N-$p$ bands stretching down to about 7 eV below the VBM and conduction band energies up to about 7 eV above the VBM. This is adequate to converge the BSE evaluation of $W({\bf q})$.

The optical dielectric function is calculated using both the \textit{independent (quasi)particle approximation} (IPA) and the Bethe Salpeter equation in the Tamm-Damcoff approximation and with static $W$ \cite{Onida02} as implemented by Cunningham \etal \cite{Cunningham23} in LMTO basis set within the {\sc Questaal} suite. 
In the IPA, we find the imaginary part of the dielectric function, $\varepsilon_2$, by calculating the velocity matrix elements for direct transitions from the valence states $\psi_{v{\bf k}}$ to conduction states $\psi_{c{\bf k}}$ with energies $\epsilon_{v{\bf k}}$, $\epsilon_{c{\bf k}}$ in $GW$:

\begin{equation}
\begin {split}
	\varepsilon_2(\omega)&=\frac{8\pi^2}{\Omega \omega^2} \sum_{cv{\bf k}}(f_{v{\bf k}}-f_{c{\bf k}})\left|\bra{\psi_{v{\bf k}}}\hat{\bm{e}}\cdot \bm{v}\ket{\psi_{c{\bf k}}}\right|^2 \\ &\times\delta(\epsilon_{c{\bf k}}-\epsilon_{v{\bf k}}-\omega),
\end{split}
\label{eps2}
\end{equation}
where $\Omega$ is the unit cell volume, $\hat{\bm{e}}$ the polarization, $\bm{v}$ the velocity operator, and $f_{n{\bf k}}$ are Fermi occupation factors for the conduction bands $c$ and valence bands $v$. The advantage of this approach is that we can carry out the Brillouin zone integrations using the tetrahedron method and use a well-converged {\bf k}-mesh. Also, it is possible to take the
sum over bands apart into contributions from band pairs.

In the BSE in the Tamm-Dancoff approximation \cite{Hanke78,Strinati1988,Rohlfing2000,Onida02}, the dielectric function is similarly calculated from the eigenvalues and eigenvectors of a two-particle Hamiltonian,
\begin{eqnarray}
	H^{(2p)}_{n_1n_2{\bf k},n_1^\prime n_2^\prime{\bf k}^\prime}({\bf q})&=&\left(\epsilon_{n_2{\bf k}+{\bf q}}-\epsilon_{n_1{\bf k}}\right)\delta_{n_1n_1^\prime}\delta_{n_2n_2^\prime}\delta_{{\bf kk}^\prime}\nonumber\\
	&&-\left(f_{n_2{\bf k}+{\bf q}}-f_{n_1{\bf k}}\right) K_{n_1n_2{\bf k},n_1^\prime n_2^\prime{\bf k}^\prime}({\bf q})\nonumber, \\
\end{eqnarray}
with $f_{n{\bf k}}$ the Fermi occupation function for band $n$ at ${\bf k}$,
which includes the electron-hole interactions kernel $K_{n_1n_2{\bf k},n_1^\prime n_2^\prime{\bf k}^\prime}({\bf q})$ expanded in the basis of one-particle eigenstates $\psi_{n\bf k}(\bf r)$. This effectively mixes the direct vertical band to band transitions $(\epsilon_{c{\bf k}}-\epsilon_{v{\bf k}})$ at different {\bf k} into new ``exciton'' eigenstates $E^\lambda$ \cite{mngep2}. 
Diagonalizing this Hamiltonian in the Tamm-Dancoff approximation, where $n_1$ is restricted to be a valence state and $n_2$ a conduction band state,
one obtains the exciton eigenvalues $E^\lambda({\bf q})$ and eigenvectors
$A^\lambda_{n_1n_2{\bf k}}({\bf q})$, \ie $A^\lambda_{vc{\bf k}}({\bf q})$. Furthermore, excitonic band weight contributions, $W_{c\bf k}=\sum_{v}\left|A^{\lambda}_{vc\bf k}({\bf q})\right|^2$ and $W_{v\bf k}=\sum_{c}\left|A^{\lambda}_{vc\bf k}({\bf q})\right|^2$, for a given conduction $c\bf k$ and valence $v\bf k$ band of the two-particle states in a narrow energy range $E_{\rm min} \le E^{\lambda}({\bf q}) \le E_{\rm max}$, can be obtained from the exciton eigenvectors. Here $\lambda$ denotes the excitonic eigenstate. 

In addition, the macroscopic dielectric function is then given by
\begin{eqnarray}
	\varepsilon_M(\omega)&=&1-\lim_{{\bf q}\rightarrow0}\frac{8\pi}{|{\bf q}|^2\Omega N_k}
	\sum_{ss^\prime}(f_{n_2^\prime{\bf k}^\prime+{\bf q}}-f_{n_1^\prime{\bf k}^\prime})\nonumber \\
	&&\rho_s({\bf q}) \sum_\lambda\frac{A^\lambda_s({\bf q})A^{\lambda*}_s({\bf q})}{E^\lambda({\bf q})-\omega\pm i\eta} \rho_{s^\prime}({\bf q})^*,
	\label{epsmac}
\end{eqnarray}
introducing the shorthand $s=\{n_1n_2{\bf k}\}$, with the matrix elements,
\begin{equation}
	\rho_{n_1n_2{\bf k}}({\bf q})=\langle\psi_{n_2{\bf k}+{\bf q}}|e^{i{\bf q}\cdot{\bf r}}|\psi_{n_1{\bf k}}\rangle.
\end{equation}
Here, we assumed a non-spin polarized case, with a factor two included for spin and $N_k$ is the number of {\bf k}-points in the Brillouin zone.

We note that neither of these methods include indirect band to band transitions assisted by electron-phonon coupling or the corrections due to zero-point-motion. These are estimated to be of order -0.1 to -0.2 eV \cite{Punya11,Cardona05}. In the present case, we notice that the lowest gap is direct so the neglect of indirect transitions is of no concern. 

Finally, the effective mass tensor elements, $M^{-1}_{\alpha\beta}$ can be calculated from the one-particle states $\ket{\psi_{n\bf k}}$, $\epsilon_{n\bf k}$, within the $\bf k\cdot p$ theory:
\begin{equation}
	\begin{split}
	&(\bm{M}^{-1})_{\alpha\beta}=\delta_{\alpha\beta}\frac{1}{m_e}+\frac{1}{m_e^{2}} \times \\ &\sum_{n_2\ne n_1}\frac{\bra{\psi_{n_1\bf k}}\hat{\bm{p}}_{\alpha}\ket{\psi_{n_2\bf k}}\bra{\psi_{n_2\bf k}}\hat{\bm{p}}_{\beta}\ket{\psi_{n_1\bf k}} + \rm{c.c.}}{\epsilon_{n_1\bf k}-\epsilon_{n_2\bf k}},
	\end{split}
\label{eff-mass}	
\end{equation}
where $m_e$ is the electron mass, $\hat{\bm{p}}_{\alpha,\beta}$ are components of the momentum operator, and $\rm{c.c.}$ stands for the complex conjugate of the preceding terms. 
%Details of the convergence parameters used are as follows \textcolor{red}{include the parameters used in IPA and BSE optics: 20x20x20 mesh used in IPA. 5x5x5 mesh with eta=0.005 and nv=15 and nc=13 used in BSE}.
        
\section{Results}
\subsection{Crystal structure and stability}
The crystal structure is shown in Fig.\ref{structure}. It belongs to the space-group number 33, $Pna2_1$ or $C^9_{2v}$. We use the symmetrized primitive cell containing 16 atoms obtained from the Materials Project (MP) \cite{Jain2013}, item mp-1029633 \cite{mp-1029633}. The structural lattice parameters are $a=6.124 \textup{~\AA}$, $b=7.719 \textup{~\AA}$, and $c=5.619 \textup{~\AA}$. Each type of atom sits at the $4a$ Wyckoff position in the primitive unit cell with
%$(x,y,z)_{\rm Ca}=(0.08296,0.62269,0.97745)$, $(x,y,z)_{\rm Sn}=(0.06941,0.12721,0.99422)$, $(x,y,z)_{\rm N_{Ca}}=(0.04805,0.10296,0.36161)$, and $(x,y,z)_{\rm N_{Sn}}=(0.10033,0.64742,0.40772)$ 
$(x,y,z)_{\rm Ca}=(0.083,0.623,0.977)$, $(x,y,z)_{\rm Sn}=(0.069,0.127,0.994)$, $(x,y,z)_{\rm N_{Ca}}=(0.048,0.103,0.362)$, and $(x,y,z)_{\rm N_{Sn}}=(0.100,0.647,0.408)$
in reduced coordinates.

We note that the ratio $a/b=0.793$ is significantly smaller than the ideal $\sqrt{3}/2=0.866$ of the wurtzite supercell. We can define the equivalent wurtzite lattice constant by the area of the unit cell in the basal plane $a_w^22\sqrt{3}=ab$, which gives $a_w=3.693$ \AA. The wurtzite type $c/a_w$ is then 1.521 which is much smaller than the ideal ratio of $\sqrt{8/3}\approx1.633$. In fact it is even lower than that of AlN, where the $c/a$ is 1.602.
There is thus a strong distortion due to the large difference in ionic radii of Sn (Shannon radius 0.55 for four-fold coordination) and Ca (Shannon radius 1 for six-fold coordination). As we will show later, this has an important impact for the crystal field splitting of the valence band maximum.

The lattice constants of CaSnN$_2$ are much larger than GaN or 4H-SiC and these materials would therefore not be suitable substrates for epitaxial growth. The $a$-lattice constants of GaN and 4H-SiC are 3.189 \AA\ and 3.073 \AA\ respectively. The effective wurtzite lattice constant of CaSnN$_2$ of 3.693 \AA\ being 16\% (20\%) larger would put CaSnN$_2$ under significant biaxial compressive strain on the basal plane of these substrates.
More specifically in the  $a$-direction, the mismatch of CaSnN$_2$ to GaN would be 11\% and in the $b$-direction it would be 21\%. While this biaxial compression would lead to a larger $c/a$ which could reverse the crystal field splitting, it is probably too large to realize epitaxial growth. This is related to the larger ionic sizes of both Ca and Sn. A better match would result with InN which has lattice constant 3.53 \AA, and thus would give a mismatch of 4.6\% in wurtzite lattice constant or 0.1\% in $a$- and 9\% in the $b$-direction. However, bulk or thick film InN is not widely available.  Other substrates could potentially be chosen, such as sapphire. In particular, r-plane sapphire has previously been successfully used \cite{Mintairov2000} as substrate for growth of another II-IV-N$_2$ semiconductor, ZnSiN$_2$ and was in that case shown to lead to epitaxy with the $(010)$ plane of the $Pna2_1$ structure parallel to the  $(1\bar{1}02)$ plane of sapphire. Sapphire Al$_2$O$_3$ has lattice constants of $a=4.765$ \AA, $c=12.982$ \AA, and in the r-plane, the lattice constants are 4.765 and 7.693 \AA. This suggests that $b$ and $c$ of CaSnN$_2$ would have mismatch of $7.719/7.693$ or 0.3\% and $5.619/4.765$ or 18\% respectively if the $(100)$ plane of CaSnN$_2$ is parallel with the r-plane and $[001]$ of CaSnN$_2$ is parallel to $[10\bar{1}0]$ of sapphire. This would entail that $5c({\rm CaSnN}_2)\approx6a({\rm Al}_2{\rm O}_3$ or a small rational match with dislocations every 5 lattice constants. Another suitable substrate might be SrO (111) which would have an in-plane lattice constant of 3.63 \AA. This would amount to a more reasonable 1.7\% compressive strain.

%\textcolor{red}{Additionally, our DFT calculation in the PBEsol gives a cohesive energy (energy relative to that of the free atoms) of 5.687 eV per atom for CaSnN$_2$, 2.05 eV per Ca in the fcc structure, 4.138 eV per Sn atom in the diamond structure and 10.49 eV per N$_2$ molecule (in PBE). This yields a formation energy of $-1.517$ eV for CaSnN$_2$ relative to its elements in their respective ground state phases.}
Additionally, the materials project reports a formation energy of $-$0.617 per atom. 
As competing binary phases, Sn$_3$N$_4$ is reported to have a positive energy of formation of 0.052 eV/atom and Ca$_3$N$_2$, the most stable Ca-N compound, has a formation energy of  $-0.96$ eV per atom. Thus the reaction
\begin{equation}
  \mathrm{Ca}_3\mathrm{N}_2+3\mathrm{Sn}+2\mathrm{N}_2 \rightarrow 3\mathrm{CaSnN}_2,
\end{equation}
is exothermic by 2.56 eV, showing that CaSnN$_2$ is stable with respect to
Ca$_3$N$_2$, Sn and N$_2$.
We have also performed phonon calculations, which do not show any imaginary phonon frequencies. This indicates CaSnN$_2$ is also dynamically stable. The phonon calculations and related spectra will be reported in a future paper. 

\subsection{Band Structures}

\begin{table}[h]
	\centering
	\caption{Band gaps within various computational levels at the $\Gamma$ point in $Pna2_1$ CaSnN$_2$.}
	\begin{ruledtabular}
		\begin{tabular}{ccccc}
			transition & \multicolumn{4}{c}{gap (eV)}\\ \hline
			&GGA&$G^0W^0$&QS$GW^{\rm RPA}$&QS$GW^{\rm BSE}$\\ 
			$\Gamma-\Gamma$ &1.352&2.438&2.804&2.593 \\			
		\end{tabular}
	\end{ruledtabular}
	\label{bands}        
\end{table}

\begin{figure}
	\includegraphics[width=8cm]{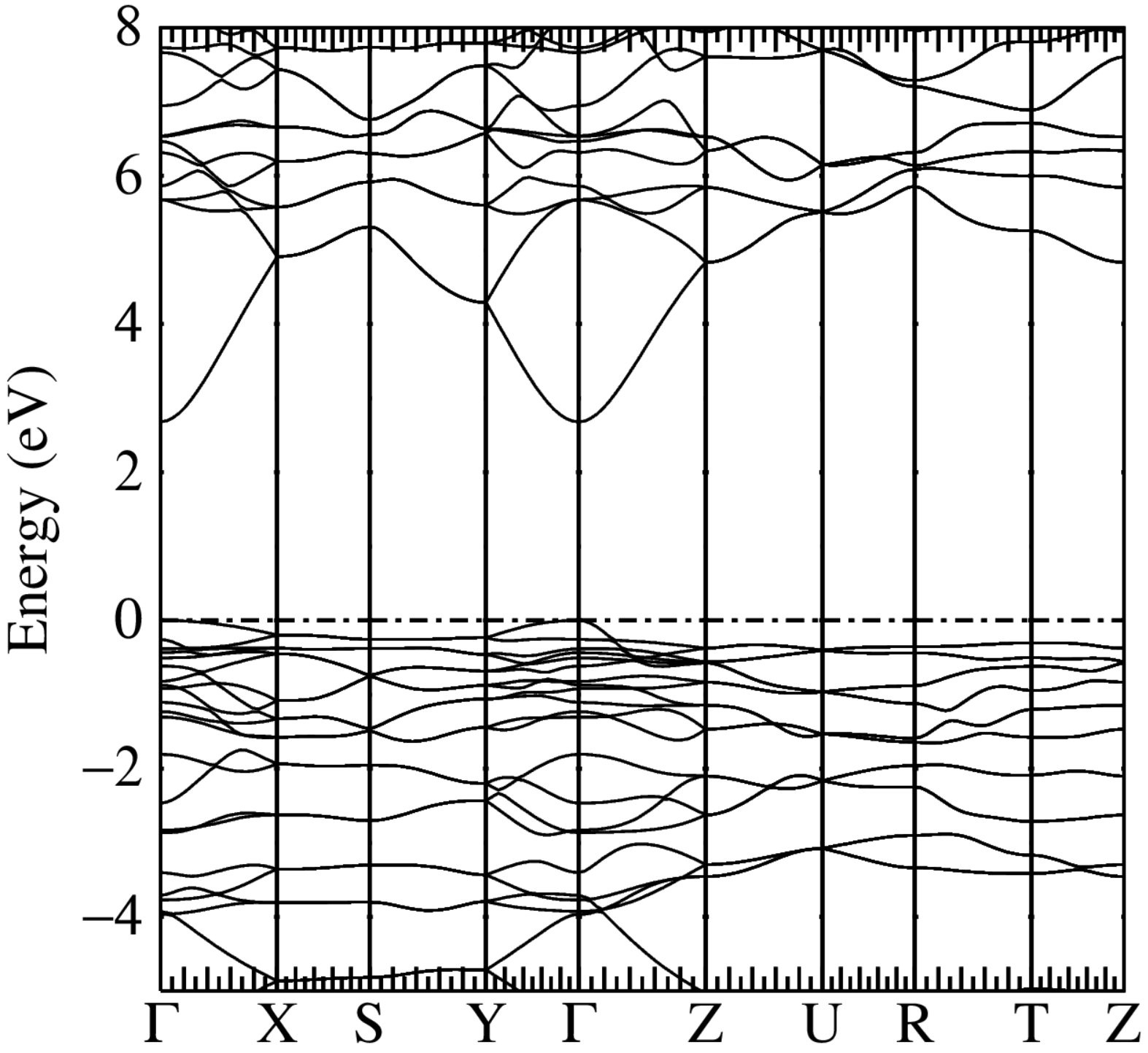}
	\caption{Band structure of CaSnN$_2$ obtained in the QS$GW^{\mathrm{BSE}}$ method, the irreducible Brillouin zone is a rectangular box with corners
          $\Gamma=(0,0,0)$, $X=(\pi/a,0,0)$, $Y=(0,\pi/b,0)$, $S=(\pi/a,\pi/b,0)$, $Z=(0,0,\pi/c)$, $U=(\pi/a,0,\pi/c)$, $T=(0,\pi/b,\pi/c)$, $R=(\pi/a,\pi/b,\pi/c)$.
	\label{fig-bnds}}
\end{figure}

The band structure in the QS$GW^{\rm BSE}$ approach is shown in Fig. \ref{fig-bnds}. The band gap in the QS$GW^{\rm BSE}$ is 2.593 eV which corresponds to $\lambda=hc/E_g$ of 478 nm in the blue region of the visible spectrum. 
The band gaps in different computational approximations are given in Table \ref {bands}. We can see that the gap in QS$GW^{\rm RPA}$ is somewhat larger than in
QS$GW^{\rm BSE}$, while the $G^0W^0$ gap, which gives the first iteration result after the GGA, is smaller but still significantly larger than the GGA. Note that the $G^0W^0$ already includes off-diagonal elements of the $\Sigma$ self-energy and thus differs from the usual single-shot perturbative $G^0W^0$. 
For the QS$GW^{\rm RPA}$ we used either a $3\times3\times3$ mesh which give 2.804 eV or a $5\times5\times5$ mesh which gave 2.776 eV, thus the $3\times3\times3$ overestimates the gap by only 28 meV.
For the QS$GW^{\rm BSE}$ case we used 10 valence and 8 conduction bands with the $5\times5\times5$ mesh, which gave a gap of 2.680 eV, or, more accurately, a $3\times3\times3$ mesh with 24 valence and 12 conduction bands, which gave a gap of 2.592 eV. In this case, all N-$2p$ derived bands are included as valence bands and conduction bands up to about 6 eV.
The gap correction from GGA to QS$GW^{\rm RPA}$ is reduced by a factor 0.85 when going to QS$GW^{\rm BSE}$. This correction is somewhat sensitive to the number of bands included in the BSE treatment of the $W^{\rm BSE}({\bf q},\omega)$. When using the smaller number of bands, it would amount to a factor 0.92. Including even more conduction bands could reduce it further. If we 
re-scale the self-energy of QS$GW^{\rm RPA}$ by the usual 0.8 reduction factor, we obtain a gap of 2.49 eV, which still corresponds to the blue-green region. Furthermore. there remains some uncertainty of order 0.1 eV on this gap because of the zero-point motion correction which would lower the gap and because the PBE lattice constant may be slightly overestimated which could lead to an underestimate of the gap. So these last two corrections would be expected to partially cancel each other. 

\begin{figure}
	\includegraphics[width=8cm]{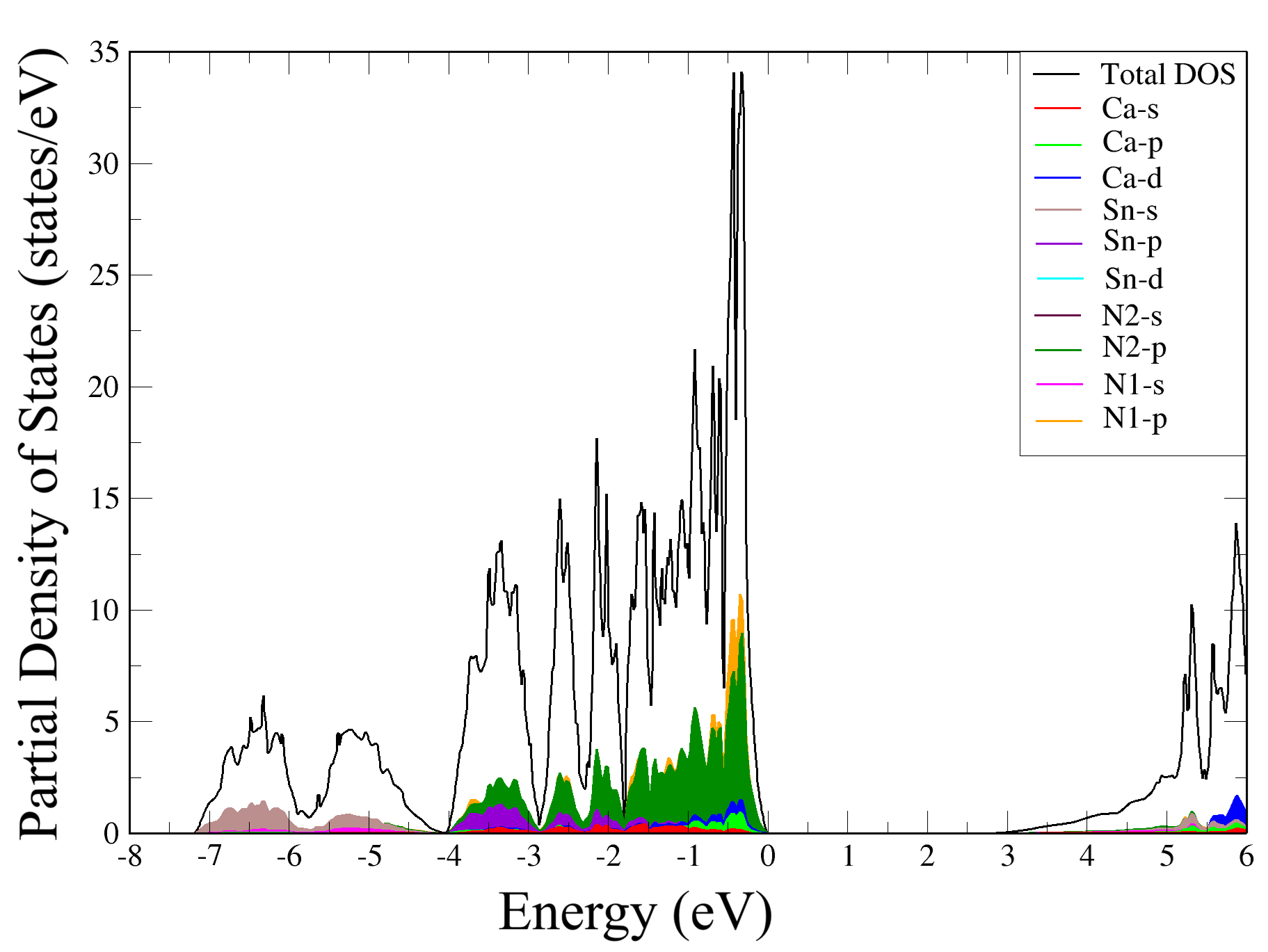}
	\caption{Total and partial densities of states in $Pna2_1$ CaSnN$_2$. The partial contributions include the sum over all equivalent atoms, but refer to partial wave contributions inside the muffin-tin spheres only, excluding those from the interstitial region and only showing the major contributions. Here N$_1$ are the Nitrogen labeled as $\rm N_{Ca}$, and N$_2$ are $\rm N_{Sn}$ in Fig. \ref{structure}. 
	\label{fig-dos}}
\end{figure}

In Fig. \ref{fig-bnds}, we can see that the minimum gap is direct at $\Gamma$. The conduction band minimum (CBM) shows a strong dispersion, characteristic of a small effective mass.
The valence band maximum (VBM) occurs at $\Gamma$ and has its strongest dispersion along $\Gamma Z$. As explained in \cite{Dernek2024}, this indicates that the VBM has $a_1$ as irreducible symmetry representation, because $z$ corresponds to $a_1$. Since the conduction band minimum is $s$ like in character and also of $a_1$ symmetry, the transitions from the top VBM to the CBM are allowed for ${\bf E}\parallel {\bf c}$ or $z$-polarization. This is similar to AlN and is somewhat disadvantageous for normal incidence light extraction for basal plane oriented films but not if the films would have the $c$-plane in the plane of the film. This orientation is actually found to occur for growth of other II-IV-N$_2$, for example ZnSiN$_2$ \cite{Mintairov2000}, on $r$-plane sapphire. The next valence band separated by a crystal field splitting of 0.257 eV  has its smallest mass along $\Gamma X$ and is thus of $x$ or $b_1$ symmetry and has allowed transitions to the CBM for ${\bf  E}\parallel{\bf a}$. We confirmed these irreducible symmetry labeling by inspection of the eigenvectors. We give the symmetry labels and values relative to the VBM of a few valence and conduction bands near the gap in Table \ref{tablevels}. We can see that only the VBM$-$4 has the $b_2$ or $y$-like character and two other states, one with $a_2$ and another one with $b_1$ symmetry lie in between. In the conduction band, the next two higher levels at $\Gamma$ are very close to each other. 
\begin{table}
  \caption{Symmetry labeled energy levels at $\Gamma$ relative to VBM within QS$GW^{\rm BSE}$ for the bottom three conduction bands and top six valence bands.}
    \label{tablevels}
  \begin{ruledtabular}
    \begin{tabular}{ccc}
      \multicolumn{1}{c}{Level} & \multicolumn{1}{c}{Symmetry} & \multicolumn{1}{c}{Energy (eV)} \\ \hline
      CBM+2 &$b_2$ & 5.680\\  
      CBM+1 &$b_1$ & 5.678\\
      CBM &$a_1$ & 2.680 \\\hline
      VBM &$a_1$ & 0      \\
      VBM-1 &$b_1$ & -0.257 \\
      VBM-2 &$a_2$ & -0.380 \\
      VBM-3 &$b_1$ & -0.434 \\ 
      VBM-4 &$b_2$ & -0.505 \\
      VBM-5 &$a_1$ & -0.615 \\  
    \end{tabular}
  \end{ruledtabular}
\end{table}

The effective mass tensor components are given in Table \ref{tabmass}. Because of the orthorhombic symmetry, the effective mass tensor is diagonal but with different values in each of the Cartesian directions. We can see that the CBM has almost isotropic symmetry, while the valence bands have quite different masses in different directions. Note that the VBM$-$1 has one of its mass tensor components of positive sign and two of negative sign, meaning that it is a saddle point instead of a maximum. The VBM$-$2 corresponding to $a_2$ symmetry has very flat band and hence high effective hole mass. The VBM$-$4 has its smallest negative mass along the $y$ direction, consistent with its $b_2$ symmetry. The CBM$+$1 and CBM$+$2 again are saddle point like as one can also directly see in the band structure
Fig.\ref{fig-bnds}. The effective mass tensor can qualitatively be understood
from Eq. \ref{eff-mass}. The closer the energy $\epsilon_{n_2\bf k}$ of band $n_2\bf k$ is to the energy $\epsilon_{n_1\bf k}$ of band $n_1\bf k$, the larger the deviation of the inverse effective mass from $1/m_e$ will be. A nearby band of higher energy tends to make the effective mass negative, while a nearby band of lower energy tends to make the effective mass positive. On the other hand, as an example, the top valence band of $a_1$ symmetry only can couple to the higher lying $a_1$ symmetry for the $p_z$ operator in the matrix elements and this explains why the largest negative dispersion occurs along $\Gamma Z$, likewise for the next VBM$-$1 of $b_1$ character. 
The momentum matrix elements are larger between the valence bands and
the lowest conduction band than to lower lying valence bands  because of the atomic orbital character of the bands involved. The coupling to the higher conduction bands plays less of a role because for these the energy denominators in Eq.\ref{eff-mass} become large.

\begin{table}
  \caption{Effective mass tensor components in atomic units for the bottom three conduction bands and top six valence bands at the $\Gamma$ point.}\label{tabmass}
  \begin{ruledtabular}
    \begin{tabular}{lccc}
    Level & $M_{xx}$ & $M_{yy}$ & $M_{zz}$ \\ \hline
    CBM+2 &-0.248 &-0.434 & 1.525 \\  
    CBM+1 &-0.785 & 6.468 &-0.278 \\
    CBM   & 0.224 & 0.208 & 0.252 \\ \hline
    VBM   &-5.824 &-3.010 &-0.270 \\
    VBM-1 &-0.331 & 2.605 &-9.810 \\
    VBM-2 &-48.710&13.843 &-0.819 \\
    VBM-3 &-4.233 & 1.934 &-0.980 \\
    VBM-4 &-4.519 &-0.884 & 8.488 \\
    VBM-5 &-2.228 &-0.675 & 2.053 \\
  \end{tabular}
   \end{ruledtabular}
\end{table}

\begin{figure}
	\includegraphics[width=8cm]{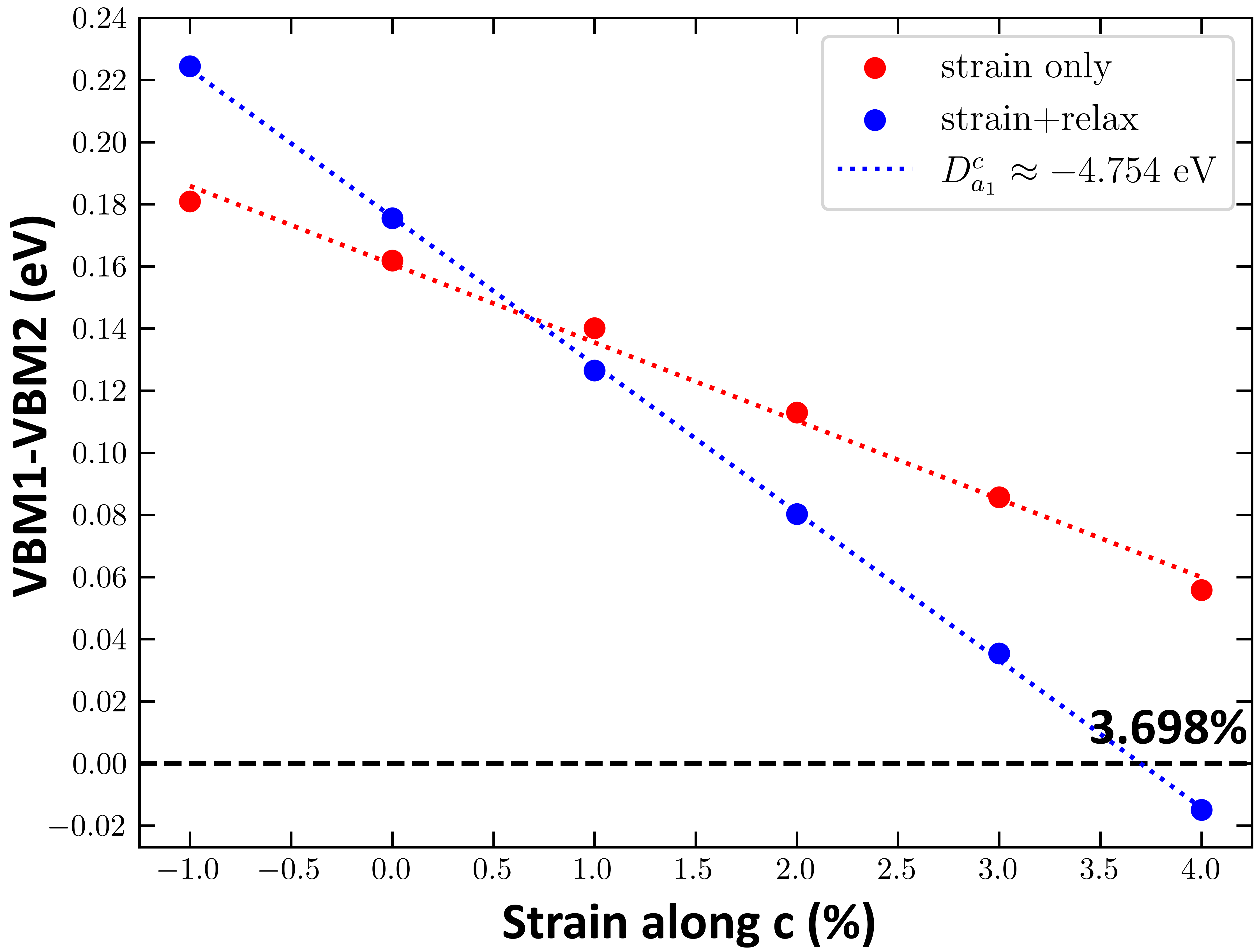} 
	\caption{Crystal field splitting between the valance band maxima (VBM1) of $a_1$ symmetry and VBM2 of $b_1$ symmetry at the $\Gamma$ point vs. various tensile strains (in percentage) applied along {\bf c} within the DFT level. Red dots show the amount of splitting under strain without any relaxation and the blue dots show the amount of splitting after the atomic positions are relaxed after the applied strain. VBM2 becomes the new maxima beyond the strain of 3.698\%. Dotted lines indicate the linear regression results where the slopes are related to the strain deformation potential $D^c_{a_1}$, which is approximately -4.754 eV for the strain+relax results.  
		\label{fig-strain}}
\end{figure}

\begin{figure*}
	\includegraphics[width=18cm]{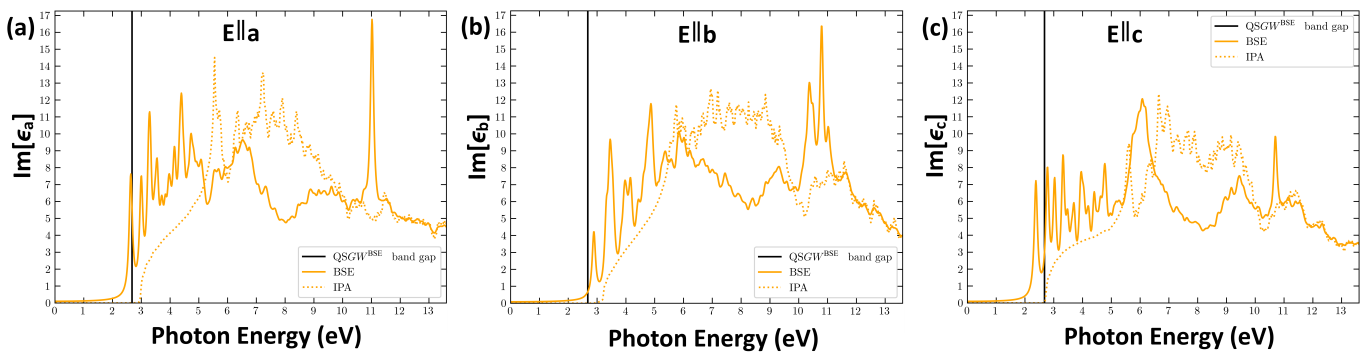}
	\caption{Imaginary part of the optical dielectric function, $\varepsilon_2^{i}(\omega)$  shown in full line, obtained using the BSE with the broadening parameter of $\eta=0.005$ Ry (as in Eq. \ref{epsmac}), including top 15 valence bands and bottom 13 conduction bands. Independent particle approximation results using a finer $20\times20\times20$ $\bf k$-point mesh, without any broadening factor (as in Eq. \ref{eps2}) is shown in dotted line. The three components of the macroscopic dielectric function parallel to $\mathbf{a}$,$\mathbf{b}$,$\mathbf{c}$ axes are shown in (a),(b),(c) respectively. 
		\label{qsgwhat_optics}}
\end{figure*}

\begin{figure*}
	\includegraphics[width=18cm]{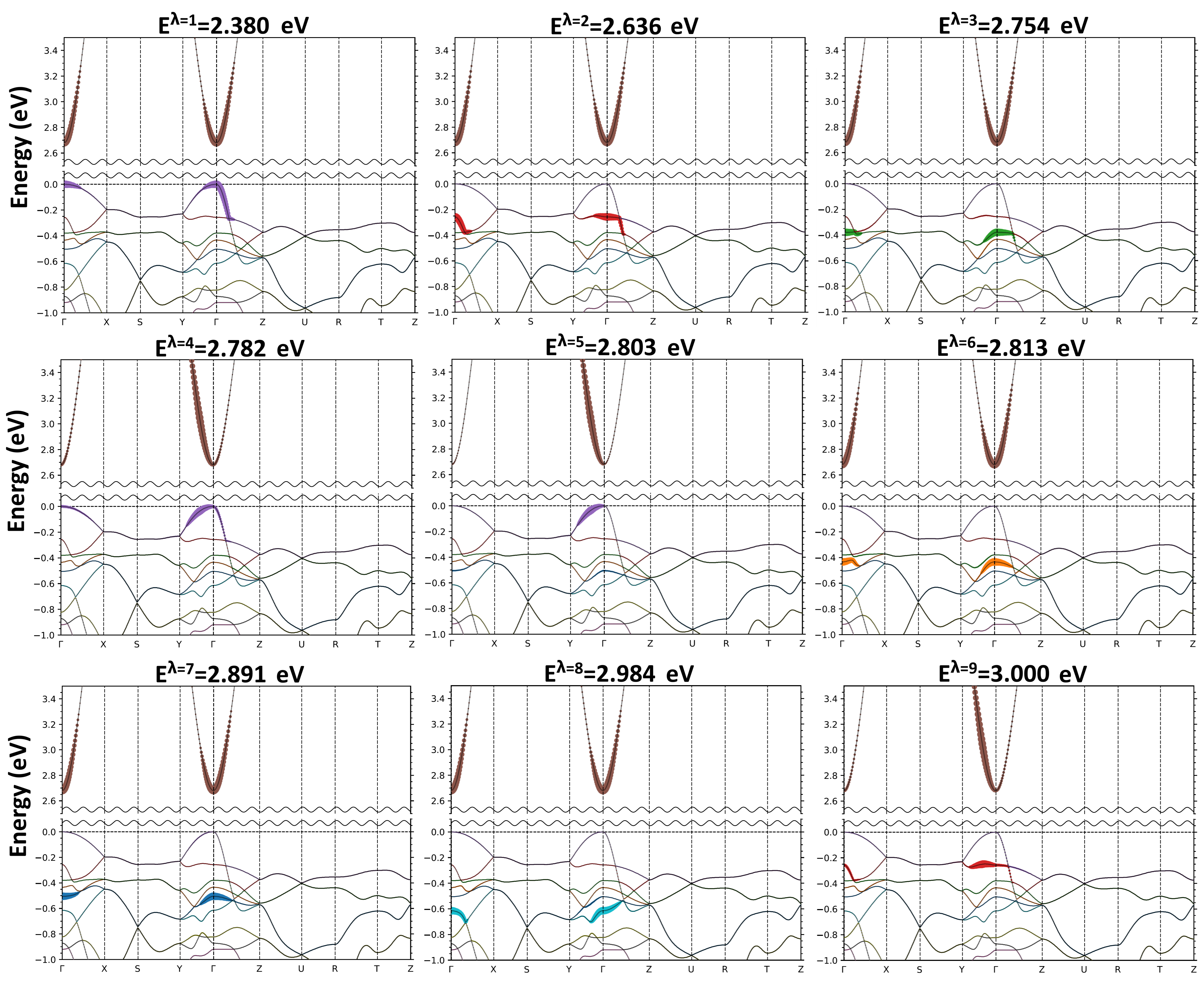}
	\caption{Exciton wave function weights, $W_{v(c)\bf k}^{\lambda}$, for $\lambda=1\dots7$ contributed by CBM of $a_1$ symmetry and top six valence bands of $a_1$, $b_1$, $a_2$, $b_1$, $b_2$, and $a_1$ symmetries. The size of the colored circles are scaled with respect to the exciton weights, and the use of different colors serves to distinguish different bands. The $\lambda=3,6,8$ excitons are dark excitons, whereas the others are bright or semi-bright with respect to their oscillator strengths reported in Table \ref{tab-exciton}.
		\label{exbw}}
\end{figure*}

The Partial Densities of States (PDOS) are given in Fig.\ref{fig-dos}. We can see that down to about 4 eV below the VBM the states have mostly N-$p$ character with both N$_1={\rm N}_{\rm Ca}$ and N$_2={\rm N}_{\rm Sn}$ giving almost equal contributions. As we move deeper into the VBM, the N-$s$ play a larger role
and hybridization with Sn and Ca increases as the states become stronger bonding in character. The analysis of the bands' atomic orbital contributions shows that the CBM is mostly Sn-$s$-like.
The Ca-$d$ states show up in the CBM at about 6 eV and as bonding states at about 0.5 eV below the VBM. 

\subsection{Strain effects}

The small $c/a$ ratio implies that the lowest optical gap is allowed for the ${\bf E}\parallel{\bf c}$ polarization, which is unfavorable for light extraction of wave vector parallel to {\bf c}. More generally, it is forbidden  for $s$-polarization (also labeled transverse electric or TE) for any incident angle on the basal plane. 
It is thus of interest to estimate how much uniaxial tensile strain is needed to reverse the valence band ordering.
To this end, we calculated the strain deformation potential defined by $D^c_{a_1}=\delta\Delta_c/\eta_{a_1}$ where the crystal field splitting, $\Delta_c=E_{a_1}-E_{b_1}$, is the energy splitting of the top two valence bands and $\eta_{a_1}$ is the strain of $a_1$ symmetry which transforms the lattice constants as $c^\prime=c(1+\eta_{a_1})$, $a^\prime=a(1-\eta_{a_1}/2)$,  $b^\prime=b(1-\eta_{a_1}/2)$. For $\eta_{a_1}>0$, $c$ is uniaxially stretched and $a$ and $b$ compressed so as to keep the strain tensor traceless or the volume unchanged.	
Fig.\ref{fig-strain} shows the amount of crystal field splitting between the $a_1$ and $b_1$ valence states at the $\Gamma$ point when a uniaxial tensile strain is applied along the {\bf c}-direction (\ie change in $c/a$ ratio) within the DFT level.  The slope is seen to be more negative if for each strain, the internal positions are relaxed instead of simply scaled.  Thus relaxation of the atomic positions in response to the strain is important. 
We find indeed that for $\eta_{a_1}$ in the range $(-0.01,0.04)$, the crystal field splitting to strain relation is linear and $D^c_{a_1}\approx-4.754$ eV. This means that the crystal field splitting would change sign for a tensile uniaxial strain of $\sim3.7$\% (\ie at $\eta_{a_1}\approx0.037$).

\subsection{Optical Properties}

\begin{figure}
	\includegraphics[width=8cm]{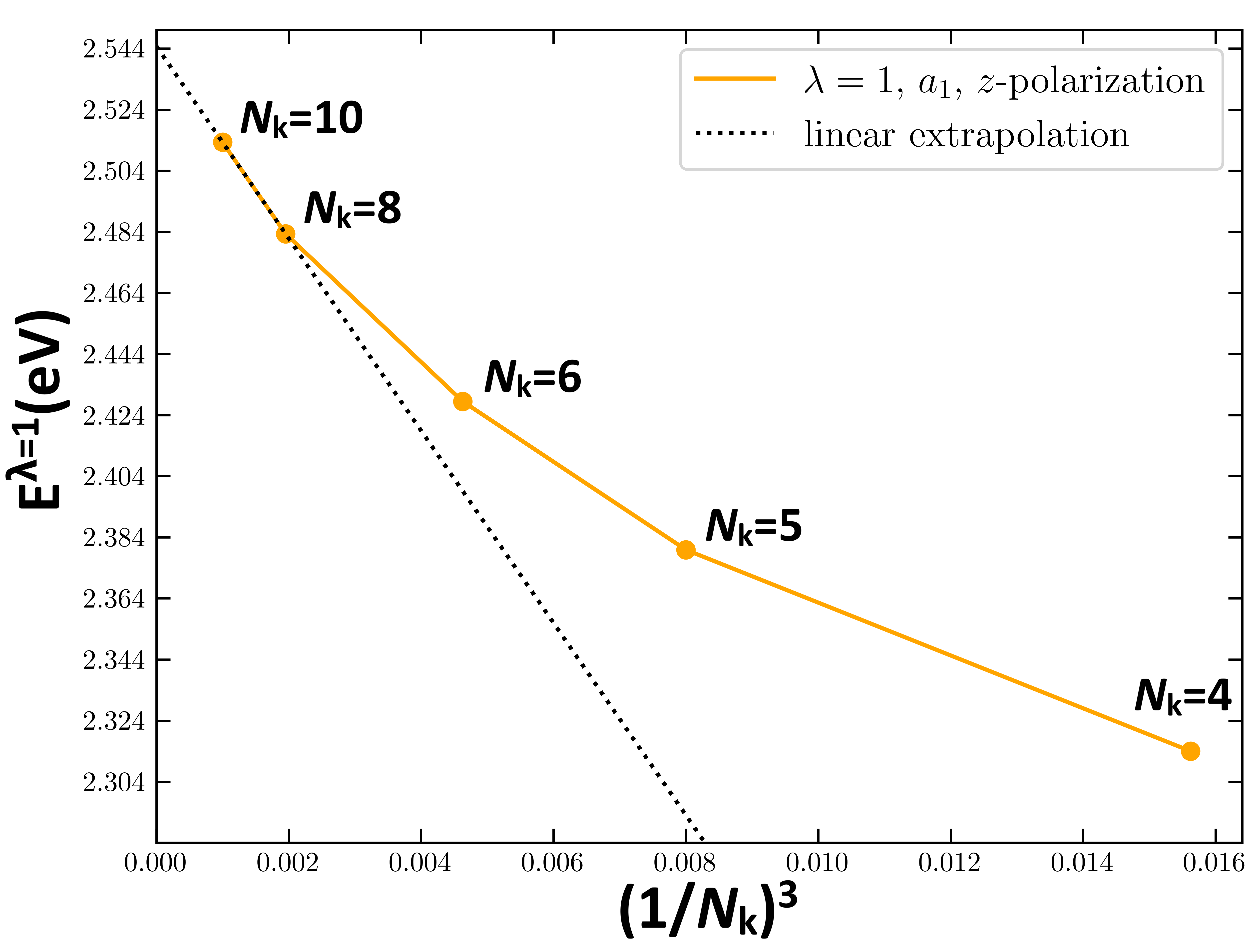}
	\caption{Exciton energy levels for $\lambda=1$, for $a_1$ symmetry with $z$-polarization, as a function of {\bf k}-mesh density. The extrapolated value is 2.545 eV. 
		\label{fig-extrapolate}}
\end{figure}

\begin{figure}
	\includegraphics[width=8cm]{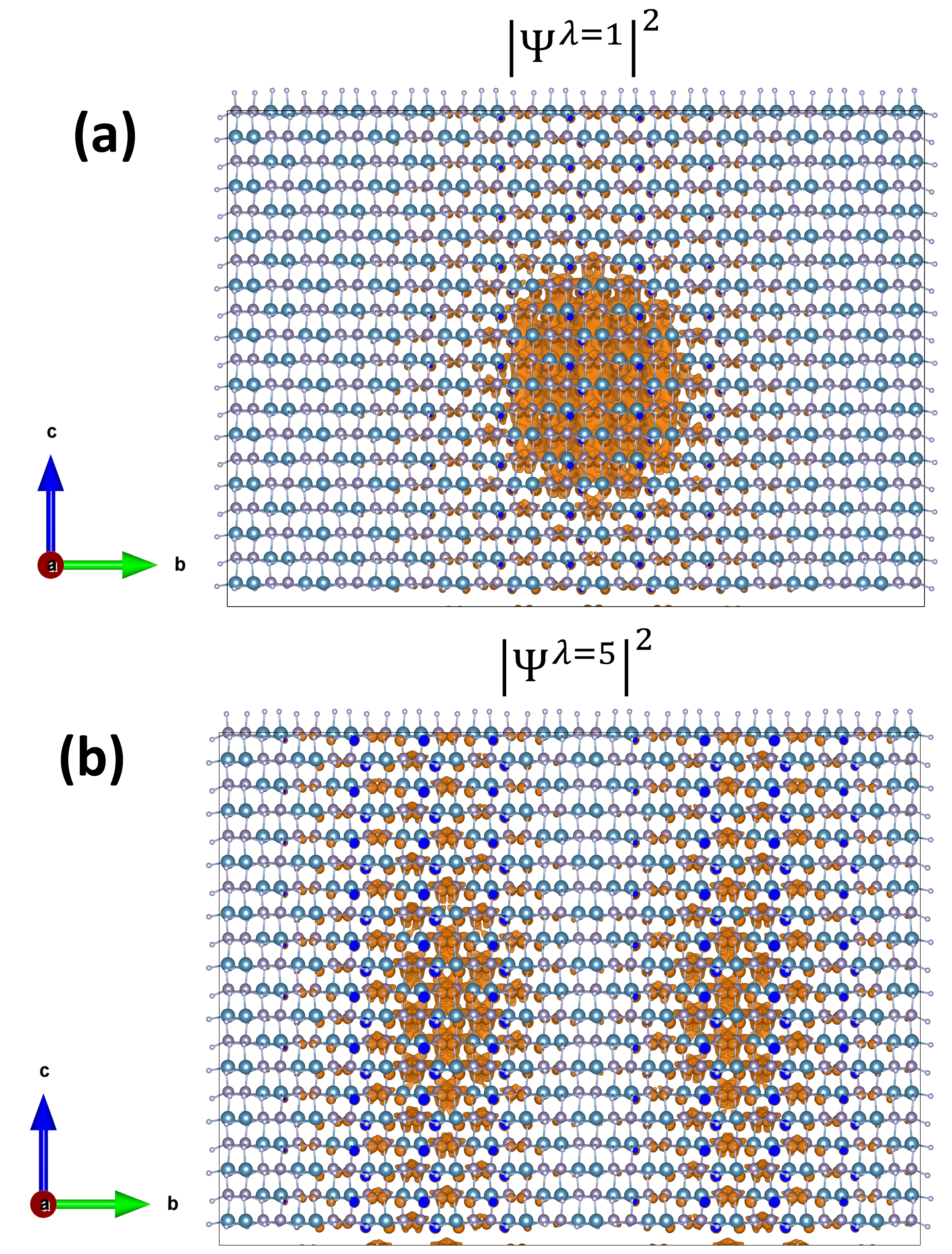}
	\caption{Position space probability density $\left|\Psi^{\lambda}\left((x,y,z)_{\rm N_{Ca}},{\bm r}_e\right)\right|^2$ of the (a) $\lambda=1$ and (b) $\lambda=5$ excitons, for a fixed hole position at $(x,y,z)_{\rm N_{Ca}}$, confined within an extended cell of $10\times10\times10$ the unit cell. Both cells are viewed from the side. The coloring of atomic sites is the same as in Fig. \ref{structure} and the probability density bubbles are shown in orange; the dark blue regions are the interior parts of the probability densities when they are cut by the supercell boundary. 
		\label{fig-ex}}
\end{figure}

%\begin{figure}
%	\includegraphics[width=7cm]{Figs/xcfh-n-2.6513-2.6515ev-10x10x10-a.png}
%	\caption{Position space probability density of the \textcolor{red}{$\lambda=5$ exciton}, for a fixed hole position at $(x,y,z)_{\rm N_{Ca}}$, confined within an extended cell of $10\times10\times10$ the unit cell. It is viewed from the side. The coloring is the same as in Fig. \ref{fig-ex1h}. 
%		\label{fig-ex5h}}
%\end{figure}

The imaginary part of the dielectric function $\varepsilon_2(\omega)$ is shown in Fig. \ref{qsgwhat_optics}.
Because of the orthorhombic symmetry, the tensor has three independent components. We show the results both in the independent particle approximation (IPA) and the Bethe Salpeter Equation (BSE) approach. The former does not include local field nor electron-hole interaction effects while the latter does include both. We here use a fairly small and energy independent broadening factor (imaginary part $\eta$ in the denominator) to show the excitonic peaks clearly below the quasiparticle gap, indicated by the vertical line. In reality we may expect that the lifetime broadening would increase with energy. We note that the results here were obtained based on the QS$GW^{\rm BSE}$ band structure obtained with the larger $5\times5\times5$ {\bf k}-mesh but smaller number of bands included. Including a larger number of bands would reduce the $W^{\rm BSE}$ and hence the gap but will then also reduce the excitonic binding energy effects, so we expect the optical gap or the lowest excitons to change only minimally. 

We can see that the lowest excitonic peak occurs for the ${\bf E}\parallel{\bf c}$ polarization, in agreement with our comments about the symmetry labeling of the corresponding bands and allowed transitions in the previous section. 
The exciton for ${\bf E}\parallel{\bf a}$ has the next higher energy and lies still below the quasiparticle gap while the exciton for ${\bf E}\parallel{\bf b}$ lies already above the quasiparticle gap. However, since this exciton is derived from a valence band below the VBM, it still has a binding energy relative to the direct gap for that polarization. The strength of the excitonic peaks is comparable to that in GaN or ZnGeN$_2$, which were studied in Ref.\onlinecite{Dernek2024}. 

The lowest few exciton eigenvalues are given in Table \ref{tab-exciton} along with their oscillator strength relative to the first bright one, their polarization and the band of corresponding symmetry from which they are derived. 
In Fig. \ref{exbw} 
the band weights $W_{v{\bf k}}$ and  $W_{c{\bf k}}$ are given as colored dots with the size of the dots indicating the weight on the band structure for some of the low lying excitons.  
We can indeed see that all the excitons shown here are derived from the conduction band minimum at $\Gamma$ but different excitons derive from different valence bands. 
The lowest exciton, which is polarized along $z$ is derived mostly from the top valence band, while the next one is derived from the VBM$-$1. The third exciton at 2.754 eV is derived from VBM$-$2 which has $a_2$ band symmetry and therefore gives a dark exciton. The fourth exciton at 2.782 eV is again derived from the top valence band, consistent with its $z$ polarization, but shows higher contribution away from $\Gamma Y$ then at $\Gamma$. This indicates it could have a radial node in k-space and real space as in a $2s$-like hydrogenic function.
This excited exciton state already lies above the corresponding band gap. 
The exciton at 2.803 eV is also derived from the top VBM but has an oscillator weight two orders of magnitude less than $\lambda=4$ exciton. Namely, this exciton has weight primarily along $\Gamma Y$ only and thus the $xz$ plane is a nodal plane. This corresponds to a $p_y$-like hydrogenic envelope function.
The $A^\lambda_{vc{\bf k}}$ for a given band pair $vc$ as function of ${\bf k}$ is the Fourier transform of a slowly varying envelope function in the hydrogenic model and can thus have different spherical harmonic character. In the present case, the hydrogenic model needs to be modified to incorporate the anisotropies resulting from the valence band symmetry and the screened Coulomb interaction.
 The fifth exciton at 2.803 eV  is $y$-polarized but with a rather small oscillator strength. Interestingly, it has a higher binding energy. 
 The exciton $\lambda=7$ at 2.891 eV is derived from VBM$-$4 and $y$- polarized.
We note that this exciton lies  
 above the nominal quasiparticle gap, indicated by the vertical lines
in Fig.\ref{qsgwhat_optics} but below the $y$-polarization onset of the IPA $\varepsilon_2$ which is at $2.680+0.505=3.185$ eV.
The exciton $\lambda=8$ at 2.984 eV is dark and is derived from VBM$-$5 of $a_1$ symmetry.

\begin{table}
  \caption{Exciton eigenvalues ($E^\lambda$), oscillator strengths ($S^\lambda$)
    relative to that of the first bright one, polarizations ($\hat e$) if bright, symmetries of the corresponding valence bands, the band energies ($\epsilon_c-\epsilon_v$), and exciton binding energies ($E_B$). All energies are given in eV.}
    \label{tab-exciton}
  \begin{ruledtabular}
    \begin{tabular}{lcccccc}
      $\lambda$&$E^\lambda$&$S^\lambda$&$\hat e$&sym&$\epsilon_c - \epsilon_v$&$E_B$\\ \hline
      1&2.380& 1   & $z$ & $a_1$ & 2.680 & 0.300 \\
      2&2.636& 1.04& $x$ & $b_1$ & 2.937 & 0.301 \\
      3&2.754& 0   & dark& $a_2$ & 3.060 & 0.306 \\
      4&2.782& 1.05& $z$ & $a_1$ & 2.680 & $-$0.102 \\
      5&2.803& 0.01& $y$ & $b_2$ & 3.185 & 0.382 \\ 
      6&2.813& 0   & dark& $a_1$ & 2.680 &$-$0.133 \\
      7&2.891& 0.56& $y$ & $b_2$ & 3.185 & 0.294 \\
      8&2.984& 0   & dark& $a_1$ & 3.295 & 0.311 \\
      9&3.000& 0.93& $x$ & $b_1$ & 3.114 & 0.114 \\
    \end{tabular}
  \end{ruledtabular}
\end{table}

One can see that the exciton binding energies measured from their corresponding band gap based on their symmetry assignment are pretty close to each other at about 0.3 eV but the dark exciton for $\lambda=6$, which is an excited exciton state corresponding to the VBM-CBM band pair with a $p$-like envelope function, has in fact a negative binding energy relative to the corresponding gap. However, the binding energies obtained here are likely to be an overestimate for two reasons. First accurate binding energies require a dense {\bf k}-point mesh near the $\Gamma$-point from which the excitons are derived in a hydrogenic approximation. We can see indeed that the excitons have band weights from a narrow range near $\Gamma$ and thus require a fine sampling near this point. Second, our calculation does not include phonon or lattice polarization effects in the dielectric screening. Typically, this will reduce the exciton binding energies by an order of magnitude \cite{Dernek2024}. This is because in the hydrogenic model the binding energy
is inversely proportional to the dielectric constant squared: the Coulomb energy is screened by the dielectric constant and the effective Bohr radius is proportional to the dielectric constant. Thus, if we replace the electronic screening only $\varepsilon_\infty$ by the static dielectric constant $\varepsilon_0$ including phonon contributions, the binding energies are divided by $(\varepsilon_0/\varepsilon_\infty)^2$. 

 The extrapolation result for the $\lambda=1$ exciton with $a_1$ symmetry can be seen in Fig. \ref{fig-extrapolate}, where we show the converged $E^{\lambda=1}$ results as a function of inverse the number of {\bf k} points in the Brillouin zone. Here we used $N_k \times N_k \times N_k$ meshes with $N_k \in$ \{4,5,6,8,10\}. We confirm that it is indeed important to converge the {\bf k}-point mesh used in the BSE two-particle Hamiltonian $H_{vc{\bf k},v'c'{\bf k}'}$ \cite{ligao2}. The extrapolated result of $E^{\lambda=1}=2.545$ eV, would correspond to a binding energy of $E_B=0.135$ eV. Additionally, we performed phonon calculations to obtain the following static and macroscopic dielectric function results from density functional perturbation theory \cite{dfpt1,dfpt2} using the {\sc Abinit} code suite \cite{abinit1,abinit2}:
 
$$
 	\varepsilon_\infty=\begin{pmatrix}
 		6.3&0&0\\
 		0&5.8&0\\
 		0&0&6.2
 	\end{pmatrix};\varepsilon_0=\begin{pmatrix}
		11.5&0&0\\
		0&10.3&0\\
		0&0&13.7 	
 	\end{pmatrix},
$$
 and from which we find the averaged values of $\varepsilon_{\infty,0}^{\rm avg.}=\left(\varepsilon_{\infty,0}^{\rm xx}\times\varepsilon_{\infty,0}^{\rm yy}\times\varepsilon_{\infty,0}^{\rm zz}\right)^{1/3}$.
 
 Thus, when we further correct the extrapolated value by dividing with a factor of $(\varepsilon_0^{\rm avg.}/\varepsilon_\infty^{\rm avg.})^2 \simeq 3.7$, we obtain a binding energy of 36.6 meV, which is close to the known value in GaN. The higher lying excitons undergo similar corrections when using a finer $k$-mesh but the relation to the band pairs and envelope function character remains the same, 

We can further check the nature of some of the excitons by examining their
wavefunction in real space. The exciton wave function is given by
\begin{equation}
\Psi^\lambda({\bf r}_h, {\bf r}_e)=\sum_{vc{\bf k}} A^\lambda_{vc{\bf k}}\psi_{v{\bf k}}({\bf r}_h)\psi_{c{\bf k}}({\bf r}_e)
\end{equation} 
and we can look at
$|\Psi^\lambda({\bf r}_h, {\bf r}_e)|^2$ as function of the electron position ${\bf r}_e$ for a chosen position of the hole ${\bf r}_h$. Since the transitions are
mostly between holes on N to the CBM, we pick the hole on a N site.
The electron distribution for the first bright exciton is shown in
Fig. \ref{fig-ex}(a) and is found to be quite spread out consistent with the shallow Wannier-type exciton. For the fifth exciton shown in Fig. \ref{fig-ex}(b) at 2.803 eV, we  can  indeed see a nodal plane consistent with the $p_y$ like envelope function mentioned earlier.

\section{Conclusions}
In this paper we showed that $Pna2_1$ CaSnN$_2$ is potentially a useful semiconductor made of sustainable and abundant elements because it has a direct  gap in the blue range of the visible spectrum. It is thus a potential alternative to In$_x$Ga$_{1-x}$N which is currently used for blue LEDs and for the technology of adding a yellow phosphorescent coating for white LEDs. The predictions of the band gap were here done with a accurate and reliable  method, the quasiparticle self-consistent QS$GW^{\rm BSE}$ method, which includes electron-hole interaction effects in the screening of $W$. We find that the small $c/a$ ratio of this compound leads to the VBM having $z$ character and hence gives smallest exciton gap  with polarization parallel to the {\bf c} axis. This is unfavorable for extraction of light from the basal plane but not if the film is grown with the {\bf c} axis in the plane of the film as would likely occur for growth on r-plane sapphire. There is a large crystal field splitting of 257 meV to the next lower valence band at $\Gamma$ which is {\bf a} polarized. However, this crystal field splitting could be reversed by a tensile uniaxial strain along the ${\bf c}$-axis of about 3.7\%. 
We presented also the effective mass tensor components for these bands, as well as a detailed analysis of the symmetry and binding energies of the excitons. Furthermore, we checked the thermodynamic and dynamical stability of the material in this crystal structure. Needless to say, much work remains to be done to understand the defect physics and doping opportunities before one can establish the full usefulness of this material for optoelectronics.

\acknowledgments{We wish to thank Professor K. Kash for bringing our attention to CaSnN$_2$ and for many invaluable conversations. This work was supported by the US Department of Energy, Basic Energy Sciences under grant number DE-SC000893 and made use of the High Performance Computing Resource in the Core Facility for Advanced Research Computing at Case Western Reserve University.}

\centerline{\bf DATA  AVAILABILITY}
The data that support the findings of this article are openly available \cite{datavail}. Embargo periods may apply.
        
	\bibliography{Bib/lmto,Bib/dft,Bib/gw,Bib/BSE,Bib/caivn2,Bib/zgn,Bib/msn,Bib/zsn}
	
\end{document}